# Car-To-Cloud Communication Traffic Analysis Based on the Common Vehicle Information Model

Johannes Pillmann, Benjamin Sliwa, Jens Schmutzler, Christoph Ide and Christian Wietfeld
TU Dortmund University, Communication Networks Institute (CNI)
Otto-Hahn-Str. 6, 44227 Dortmund, Germany
{johannes.pillmann, benjamin.sliwa, jens.schmutzler, christoph.ide, christian.wietfeld} @tu-dortmund.de

*Abstract*—Although connectivity services have been introduced already today in many of the most recent car models, the potential of vehicles serving as highly mobile sensor platform in the Internet of Things (IoT) has not been sufficiently exploited yet. The European AutoMat project has therefore defined an open Common Vehicle Information Model (CVIM) in combination with a cross-industry, cloud-based big data marketplace. Thereby, vehicle sensor data can be leveraged for the design of entirely new services even beyond traffic-related applications (such as localized weather forecasts). This paper focuses on the prediction of the achievable data rate making use of an analytical model based on empirical measurements. For an in-depth analysis, the CVIM has been integrated in a vehicle traffic simulator to produce CVIM-compliant data streams as a result of the individual behavior of each vehicle (speed, brake activity, steering activity, etc.). In a next step, a simulation of vehicle traffic in a realistically modeled, large-area street network has been used in combination with a cellular Long Term Evolution (LTE) network to determine the cumulated amount of data produced within each network cell. As a result, a new car-to-cloud communication traffic model has been derived, which quantifies the data rate of aggregated car-to-cloud data producible by vehicles depending on the current traffic situations (free flow and traffic jam). The results provide a reference for network planning and resource scheduling for car-to-cloud type services in the context of smart cities.

## I. INTRODUCTION

Modern vehicles are equipped with a strongly increasing number of complex sensors, making them highly suitable for acting as mobile sensor platform in the IoT context. Nevertheless, the use of vehicle data for non-automotive applications is not widespread as Quality-of-Service (QoS) requirements cannot be sufficiently guaranteed. The aim of this paper is to provide a model for predicting the achievable data rate in car-to-cloud vehicle sensor networks. Thus, the fulfillment of the application requirements can be evaluated in a situation-aware manner.

The car-to-cloud data traffic in this work makes use of the brand-independent Common Vehicle Information Model (CVIM) (cf. Section III). The CVIM has been developed in the European Union's Horizon 2020 project AutoMat and targets harmonization and standardization of in-vehicle sensor measurements for car-to-cloud data traffic.

For an in-depth analysis, the CVIM has been implemented into a vehicle traffic simulator to produce data streams as the result of the individual behavior of each vehicle (Section IV). The vehicle traffic has been simulated in a large scale street map of the German city Dortmund for different traffic states. The communication model, based on LTE technology, realistically imitates a large German mobile network provider. The results quantify the available data rate for car-to-cloud communication and a vehicle traffic state aware data aggregation is proposed. The analysis provides a reference for network planning and resource scheduling for car-to-cloud type services (Section V).

## II. RELATED WORK

Transferring vehicle data into the cloud for evaluation has been an ongoing research topic in the past. Different architectures and approaches were developed and evaluated, challenges and advantages have been discussed [1], [2]. In [3], a traffic state estimation in road networks using car-to-cloud communication was developed , where vehicles send their velocity data to a centralized server. In [4] and [5], the randomized transmission of Floating Car Data (FCD) was proposed in order to reduce the communication costs in traffic information sharing systems, and was evaluated in highway simulation scenarios.

The authors of [6] compared and analyzed different FCD sampling strategies and methods with the goal of reducing the amount of data. The evaluation showed a trade off between the communication costs and an error in the accuracy of positioning and speed estimation. The massive deployment of vehicular data sources and its impact on networking infrastructure was discussed and evaluated in [7]. The authors presented and proposed a job-based FCD method that allows probing at precise locations and with specific sensor types, but which is limited to a number of vehicles and thereby reduces the amount of data traffic and impact on networking infrastructure.

The collection, dissemination and multi-hop forwarding of vehicle data with LTE for car-to-car as well as car-to-infrastructure communication was simulative analyzed in [8] in terms of efficiency and packet losses. Further, relieving the impact of FCD traffic in heterogeneous and hybrid networks using LTE as well as Vehicular Ad Hoc Networks (VANETs) was presented in [9] and [10] by introducing cluster heads. [11] evaluated in addition the impact of FCD on Human to Human (H2H) data traffic. In [12], the LTE data rate for effective car-to-cloud communication was improved by applying a communication channel aware data transmission. In contrast to existing vehicular network protocols simulators [13], which use synthetic models, this work evaluates an analytical approach based on empirical measurements [14].

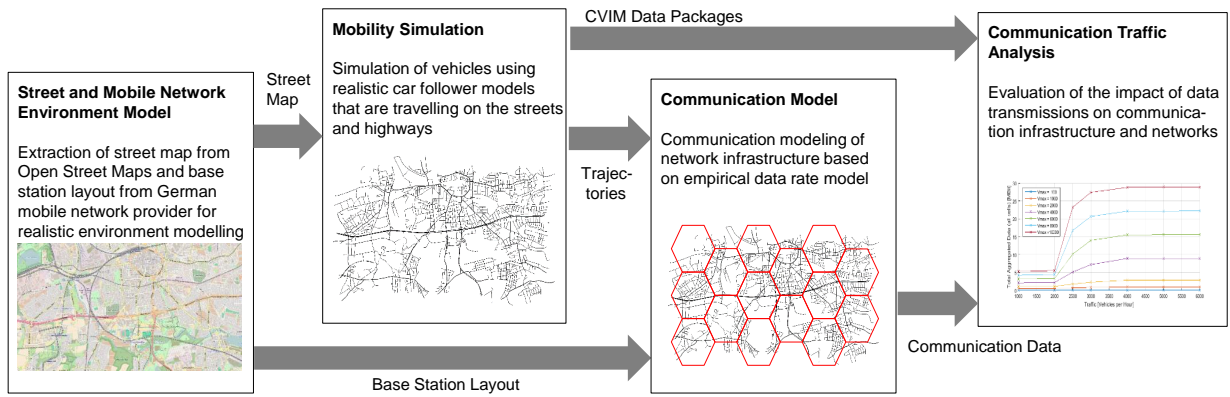

Fig. 1. Methodology of the Car-To-Cloud Communication Traffic Evaluation

## III. THE COMMON VEHICLE INFORMATION MODEL

It is today's common practice to build monolithic systems, each collecting and processing vehicle data on their own. This results in proprietary solutions that fragment the market and hinder a common ecosystem. One project approaching a holistic solution for vehicle data is the EU's Horizon 2020 AutoMat Project[1]. The project aims at creating a vehicle big data marketplace for innovative cross-sectoral vehicle data services, especially from the non-automotive industry (e.g. weather prediction services, insurance services, etc.). This marketplace requires the aggregation of preselected data without knowledge of its later intended use. The data must be measured with the highest possible quality while being in an embedded environment that is limited in computing power, communication bandwidth and data storage capacity.

One key part of this solution is a unified and efficient data representation, harmonizing proprietary data as well as anonymizing personal data and removing brand-specific information. As no standardized, non-proprietary data model exists, the novel Common Vehicle Information Model (CVIM) has been developed within the AutoMat project.

The CVIM works on three layers (cf. Figure 2). On the bottom, the signals inside the vehicles are described. These signals are proprietary and car manufacturer-specific. Their origin can be any source within the vehicle, e.g. a signal captured from Controller Area Network (CAN) or On-Board Diagnostics (OBD) bus. The next and middle layer consists of Measurement Channels. Measurement Channels define a common ground between signals of different brands. They harmonize proprietary information into a standardized format by removing brand dependencies. On the top level, aggregated data is stored inside CVIM Data Packages. These packages are exchangeable messages that can be transferred from the vehicle into the cloud. In addition, CVIM Data Packages include data ownership, copyright and privacy information, which empower the enforcement of the vehicle user's privacy rights. Certification of all transfered messages ensures completeness, validity and high quality of data.

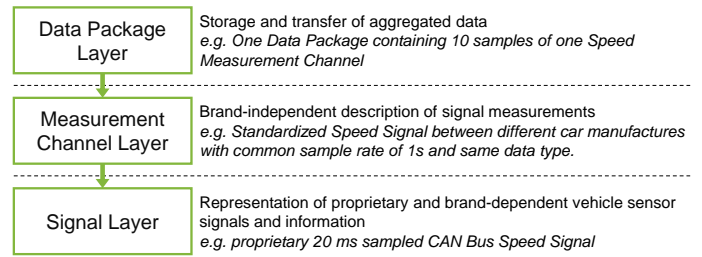

Fig. 2. CVIM High Level Architecture

## IV. CAR-TO-CLOUD COMMUNICATION SYSTEM MODEL

In this section, the simulation architecture, as shown in Figure 1, is described in detail. The results presented in this work were achieved in a four step process. First, a model of the environment was created using a road map and LTE base station locations. In the next step, the mobility of cars was simulated using the microscopic mobility simulator SUMO (Simulation of Urban MObility) [15]. Afterwards, an empirical communication model in MATLAB was performed based on the vehicle's trajectories, followed by the car-to-cloud data traffic model analysis.

### A. Street and Mobile Network Environment Modeling

The environment model is based on the map from the Open Street Map (OSM) project[2]. OSM is a community-driven project aiming at providing an open, free-to-use and highly precise map that includes streets, buildings, traffic lights, speed limits and more. In this work, the map of the city of Dortmund around its central road A40/B1 was imported and used as road map for the mobility simulation step. This scenario is characterized as very heterogeneous as it includes sections with high ($80 - 120$ $km/h$) as well as low speeds ($0 - 50$ $km/h$). In addition, different traffic states such as traffic jams and free flow can be simulated, characterized by different mean traffic densities.

The second part of the environment model provides locations of the LTE base stations and serves as input for the communication model. In this work, the base station layout of the German mobile network provider was used. These base station locations were imported from the public databases of the Federal Network Agency for Telecommunications[3] and Katasteramt Dortmund[4]. An example trajectory of a vehicle driving eastbound on the A40/B1 map with the included base station locations along the road is given in Figure 3.

---

[1]www.automat-project.eu

[2]www.openstreetmap.org

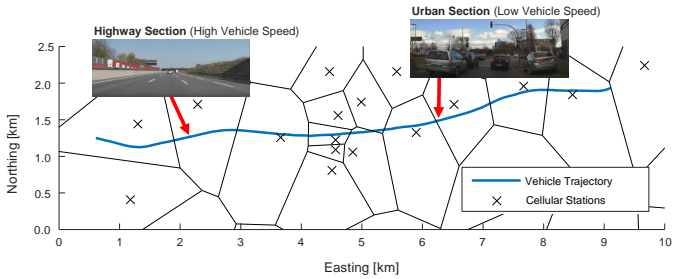

Fig. 3. Example Trajectory of one Vehicle Driving Eastbound on the A40/B1. UTM Coordinates relative to $32U\ 388000E\ 5705000\ N$.

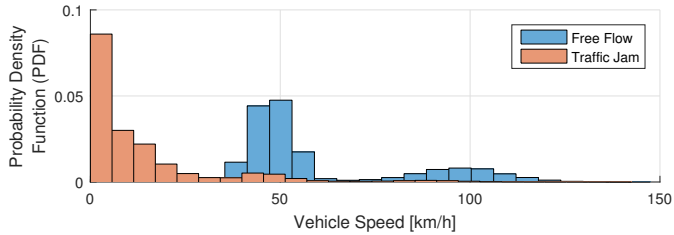

Fig. 4. Distribution of Vehicle Speeds Simulated by SUMO for Different Traffic States

### B. Vehicle Traffic Simulation

To provide reliable and realistic user mobility within this work the vehicle traffic is simulated using the open source software Simulation of Urban MObility (SUMO) [15]. SUMO is a microscopic traffic simulator where each vehicle is explicitly modeled. The simulator uses the environment road network as input. To obtain different traffic states, the traffic densities are variable. Populating the scenario with low traffic density results in a free flow traffic state and vice-versa, high traffic density leads to traffic jams. Figure 4 shows the vehicle speed distribution for both traffic states All configurations and SUMO parameters used are listed in Table I and are based on measurements by the German Federal Highway Research Institute[5].

### C. Car-To-Cloud Communication Model

Within this work, the data upload of vehicles into the cloud is realized through LTE. For this purpose, the car-to-cloud communication from the vehicle's LTE user equipment (UE) towards the LTE base station (BS) is modeled within a MATLAB simulation. Figure 5 shows the detailed architecture of the model. First, the signal quality in form of the signal-to-noise ratio (SNR) is calculated using the Winner-II B1 urban micro cell path loss model [16] in dependency of the vehicle's distance to the BS, the transmit power, noise power and figure as well as the antenna gain of the UE and BS antennas.

[3]http://emf3.bundesnetzagentur.de/karte/Default.aspx
[4]http://geoweb1.digistadtdo.de/OWSServiceProxy/client/mobilfunk.jsp
[5]http://www.bast.de/DE/Verkehrstechnik/Fachthemen/v2-verkehrszaehlung/Stundenwerte.html?nn=626916

TABLE I
SIMULATION CONFIGURATION AND PARAMETERS

| | Parameter | Value |
|---|---|---|
| **Mobility Simulation Parameters** | Simulated Vehicles | 4000 |
| | Car follower model | Krauss |
| | Lane changing model | LC2013 |
| | Maximum acceleration | $1.5\ m/s^2$ |
| | Maximum deceleration | $4.5\ m/s^2$ |
| | Maximum speed | $130\ km/h$ |
| | Speed deviation | 0.1 |
| | Vehicle length | $5.0\ m$ |
| | Driver imperfection | 0.5 |
| | Driver reaction time | $1.0\ s$ |
| | Minimum Gap | $2.5\ m$ |
| | Free flow traffic density | $1000\ Vehicles/h$ |
| | Traffic jam traffic density | $4000\ Vehicles/h$ |
| **Communication Model Parameters** | Frequency | $1.8\ GHz$ |
| | Send power | $23\ dBm$ |
| | BS / UE Antenna Gain | $15\ /\ 1\ dBi$ |
| | Noise Figure | $6\ dB$ |
| | Noise Power | $-100\ dBm$ |
| | Channel Bandwidth | $20\ MHz$ |
| | Resource Blocks per Cell | 100 |
| | LTE Scheduler | Round Robin |
| | Duplex Mode | Frequency Division Duplex |

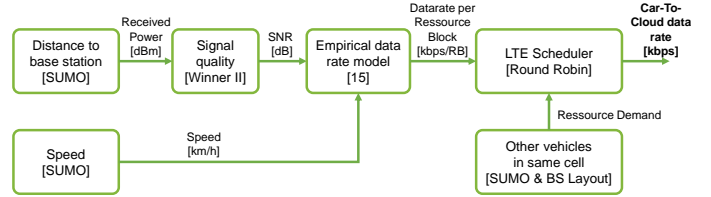

Fig. 5. Detailed Architecture of the Communication Model to Determine the Available Car-To-Cloud Data Rate

An example of the SNR is given in Figure 6. The UE is always associated to the closest base station providing the best SNR (compare Figure 3 with the vehicle's trajectory and according base stations). In the next step, the empirical data rate model from [14] is applied. The model is based on LTE uplink channel measurements and combines SNR and vehicle speed to derive the available data rate per LTE resource block (RB). Figure 7 shows the example vehicle's data rate per RB.

Subsequently, LTE transmissions are scheduled. In this work, a Round Robin (RR) scheduler was used. RR provides an equal share of RBs to all UEs within one LTE cell. Even though RR does not achieve the best performances in terms of total cell throughput, it acquires the best fairness for all vehicle's data transmissions and was therefore chosen as the best candidate for this work. Figure 8 shows the resulting data rate. The peak at $180\ s$ is caused by the fact that vehicles cross a nearly empty LTE cell. As only a few UEs are attached to this cell, each UE gets a large number of RB assigned and the data rate improves significantly up to $3\ Mbps$.

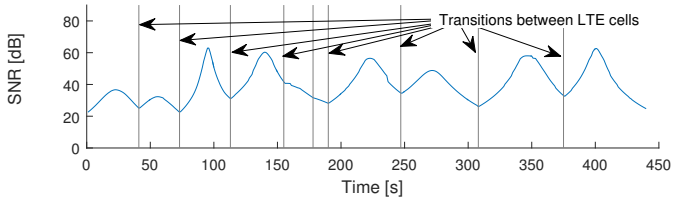

Fig. 6. Example Vehicle's Signal Quality in Form of the SNR

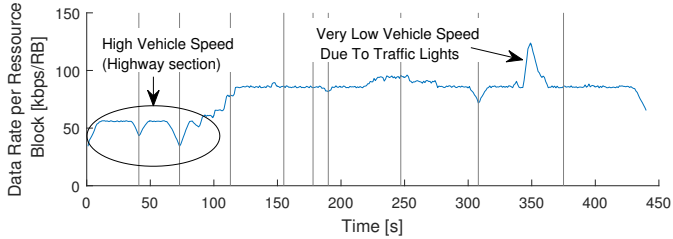

Fig. 7. Example Vehicle's Uplink Data Rate per Resource Block

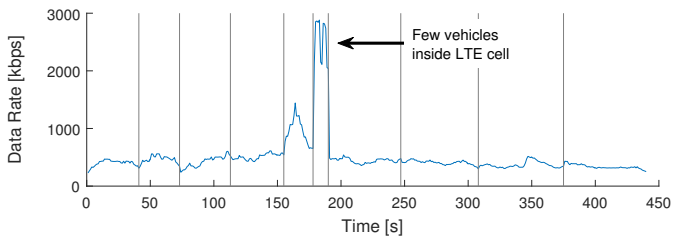

Fig. 8. Example Vehicle's Available Uplink Data Rate for Car-To-Cloud Communication

### D. Vehicular Data Traffic Model

The data sent from the vehicle into the cloud is formated in the CVIM. The data is always aggregated for one second into one CVIM data packages and then send to the cloud. There is always data available and data is only sent if the available data rate is sufficient for the application requirements.

## V. CAR-TO-CLOUD DATA TRAFFIC ANALYSIS

In this section, the analysis of car-to-cloud data traffic is performed. Figure 9 shows the available data rate that vehicles can use to send data to the cloud for the two traffic states Free Flow and Traffic Jam. The mean data rate in the Free Flow scenario is 482.1 $kbps$ and therefore seven times higher than the 69.9 $kbps$ in the Traffic Jam scenario. In the Free Flow scenario, the density of vehicles is a lot lower. The vehicles drive faster and there is a larger gap between two consecutive cars. Thus, the number of vehicles per cell is lower and each vehicle is assigned more Resource Blocks (RB) by the LTE cell. The other way around, in the Traffic Jam scenario roads are crowded by vehicles. They leave only short gaps between them and the vehicle density per road segment is higher. A larger number of cars need to share the RB which results in a lower average available data rate of 69.9 $kbps$. This data rate is sufficient to transfer various sensor data, such as as position, brakes, rain sensor, wipers, light, etc., into the cloud once per second. In the case of Free Flow, multimedia data, such as photos from in-vehicle cameras, is transferable in addition.

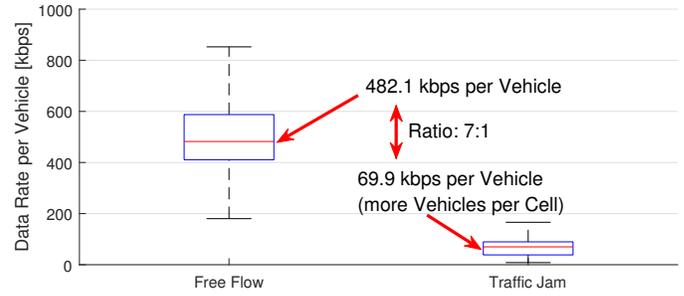

Fig. 9. Available Car-To-Cloud Data Rate per Vehicle

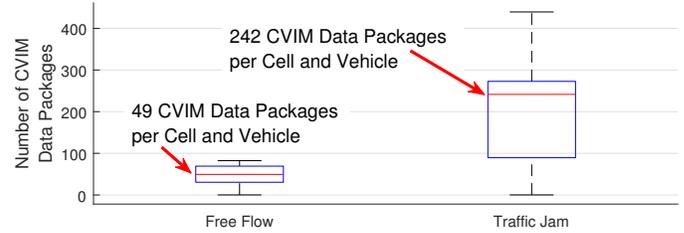

Fig. 10. Average number of generated CVIM Data Packages per Cell and Vehicle when a vehicle drives through one LTE Cell. Average over all LTE cells with vehicles traveling through.

Vehicle sensor data is aggregated in one-second intervals into CVIM Data Packages as described in the previous section. Figure 10 shows the average number of CVIM Data Packages per vehicle, that vehicles generate when driving through an LTE cell. Only LTE cells with vehicles passing through the cell are taken into account. For Free Flow, the number of packages is significantly lower than for the Traffic Jam state. In the former scenario, vehicles move faster through the LTE cells and therefore produce less packages per cell. On the other side, in traffic jams 242 CVIM Data Packages are generated in average due to the longer residence time per LTE cell. The number of generated packages behaves contrary to the available data rate. Therefore, either the number of packages or the payload per package needs to be reduced in order to match the lower available bandwidth in the Traffic Jam situation. In addition, in traffic jams vehicle sensors often detect redundant information, e.g. all vehicles report nearly the same velocity, their traffic sign recognition systems detect the same speed signs and thermometers report the same temperature. An upload is not always necessary. Therefore, the authors of this work propose traffic state aware data aggregation and car-to-cloud transfer.

The previous results were based on the assumption that the network operator provides his complete capacity in terms of RB. Usually, network operators spend only a limited amount of RBs on car-to-cloud communication. Figure 11 shows the cumulative distribution function (CDF) of the available data rate, when the number of RBs is reduced to ten. The Traffic Jam state results in 3 $kbps$ minimum data rate in 95 % of the cases and lies clearly below the Free Flow data rate with 32 *kbps*. The former case requires careful scheduling and planning of data transfers. Either the time-resolution of vehicle sensor measurements is reduced by aggregating CVIM Data Packages over longer intervals (e.g. one data package every 10 seconds), or only a selected subset of some selected, priorized data is transfered into the cloud while time-uncritical data needs to be stored for a later transfer.

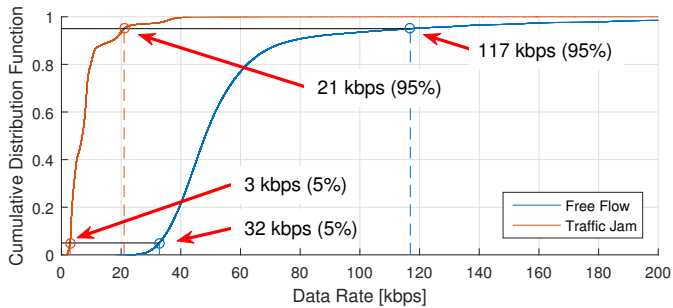

Fig. 11. CDF of the Available Data Rate per Vehicle With a Limited Number of 10 RB for Car-To-Cloud Communication

Even though the data rate is significantly reduced, the proportion between Free Flow and Traffic Jam state stays equal due to the characteristics of the Round Robin scheduler which scales linearly with the amount of RBs reserved for Car-To-Cloud communication. This relationship can also be utilized reversely. When a car manufacturer requires a guaranteed data rate for his car-to-cloud service, the model provides the amount of RBs needed.

## VI. CONCLUSION

In this work, a simulative analysis of car-to-cloud data traffic was presented. The simulation model based on a detailed environment model of the area of the German city Dortmund including a precise road map and cell locations of a large German mobile network provider. The mobility of the vehicles was simulated for different traffic states using the SUMO simulator. The car-to-cloud communication model, leveraging LTE uplink channels, was founded on a measurement-based empirical channel model. CVIM Data Packages were sent to the cloud leveraging the CVIM format.

The simulation resulted in an average data rate of 482.1 *kbps* in the case of the Free Flow traffic state and was significantly reduced by a factor of seven when traffic jams occurred. This effect occured contrary to the number of in-vehicle generated CVIM Data Packages per LTE cell that increased due to the slower vehicle speed during traffic jams. The authors of this work therefore propose an automotive traffic aware communication scheduler reducing the amount of data sent in accordance with the current vehicular traffic situation. The simulation provided an estimate for the upper and lower data rates. The results serve as a reference for network resource planning in terms of RB and resource scheduling for car-to-cloud type services.

In the future we are going to extend our work by including surrounding human-to-human data traffic into the simulation and taking more advanced LTE schedulers into account.

ACKNOWLEDGMENT

This work has been conducted within the AutoMat (Automotive Big Data Marketplace for Innovative Cross-sectorial Vehicle Data Services) project, which received funding from the European Union's Horizon 2020 (H2020) research and innovation programme under the Grant Agreement no 644657, and has been supported by Deutsche Forschungsgemeinschaft (DFG) within the Collaborative Research Center SFB 876 Providing Information by Resource-Constrained Analysis, project B4.